\documentclass[cits]{PoS}

\usepackage{eucal}
\newcommand{\be}{\begin{equation}}
\newcommand{\ee}{\end{equation}}
\newcommand{\bea}{\begin{eqnarray}}
\newcommand{\eea}{\end{eqnarray}}
\newcommand{\bse}{\begin{subequations}}
\newcommand{\ese}{\end{subequations}}
\newcommand{\sy}{\sigma_\mrm{s}}
\newcommand{\mrm}[1]{\mathrm{#1}}
\newcommand{\mc}[1]{\mathcal{#1}}
\newcommand{\blr}[1]{\left(#1\right)}

\def\Phib{\bar{\Phi}}
\def\qqb{\mrm{q\bar{q}}}
\newcommand{\pd}{\partial}

\graphicspath{{./figures/}}

\title{QCD thermodynamics of effective models with an improved Polyakov-loop potential}

\ShortTitle{QCD thermodynamics of effective models with an improved Polyakov-loop potential}

\author{
	\speaker{Rainer Stiele},$^{ab}$ Lisa M.~Haas,$^{ab}$ Jens Braun,$^{bc}$ Jan M.~Pawlowski$^{\,ab}$ and J\"urgen Schaffner-Bielich$^{\,abd}$ \vskip2ex\\
	\llap{$^a$}Institut f\"ur Theoretische Physik, Universit\"at Heidelberg,\\ Philosophenweg 16, D-69120 Heidelberg, Germany\vskip1ex\\
	\llap{$^b$}ExtreMe Matter Institute EMMI, GSI, Planckstra{\ss}e 1, D-64291 Darmstadt, Germany\vskip1ex\\
	\llap{$^c$}Institut f\"ur Kernphysik (Theoriezentrum), Technische Universit\"at Darmstadt,\\ Schlo\ss gartenstra\ss e 2, D-64289 Darmstadt, Germany\vskip1ex\\
	\llap{$^d$}Institut f\"{u}r Theoretische Physik, Goethe-Universit\"at {Frankfurt},\\ Max-von-Laue-Stra\ss e 1,  D-60438, Frankfurt am Main, Germany\vskip3ex\\	
	E-mail: \email{R.Stiele@thphys.uni-heidelberg.de} \phantom{E-mail: }\email{L.Haas@thphys.uni-heidelberg.de}, \phantom{E-mail: }\email{Jens.Braun@physik.tu-darmstadt.de}, \phantom{E-mail: }\email{J.Pawlowski@thphys.uni-heidelberg.de}, \phantom{E-mail: }\email{Schaffner@astro.uni-frankfurt.de}
}

\abstract{We analyse the role of the quark backreaction on the
  gauge-field dynamics and its impact on the Polyakov-loop potential.
  Based on our analysis we construct an improved Polyakov-loop
  potential that can be used in future model studies \cite{Haas:2013qwp}. In
  the present work,  we employe this improved
  potential in a study of a 2+1 flavour Polyakov-quark-meson model and
  show that the temperature dependence of the order parameters and
  thermodynamics is closer to full QCD.  We discuss the results for
  QCD thermodynamics and outline briefly the dependence of our results
  on the critical temperature and the parametrisation of the
  Polyakov-loop potential as well as the mass of the $\sigma$-meson.
}

\FullConference{Xth Quark Confinement and the Hadron Spectrum,\\
		October 8-12, 2012\\
		TUM Campus Garching, Munich, Germany}

\begin{document}

\section{Introduction}

In low-energy effective models for QCD such as the Polyakov-loop
extended Nambu--Jona-Lasinio (PNJL) and quark-meson (PQM) model the
gauge-field dynamics is not fully included. Specifically, the QCD
Polyakov-loop potential is approximated by its Yang-Mills (YM)
analogue.  Therefore, these effective models show some discrepancy in
comparison to QCD within functional approaches
\cite{Braun:2009gm,Pawlowski:2010ht}
and lattice simulations
\cite{Borsanyi:2010bp,Borsanyi:2010cj,Bazavov:2011nk,Bazavov:2012bp}.
To bring PNJL/PQM-type model studies closer to full QCD, the YM
Polyakov-loop potential has to be replaced by the QCD glue potential.
The latter is generated by the gauge degrees of freedom in the
presence of dynamical quarks, for details see
Ref.~\cite{Haas:2013qwp}.  We determine the change of the
Polyakov-loop potential due to the backreaction of the quarks on the
gauge degrees of freedom. We apply this improved Polyakov-loop
potential in a 2+1 flavour Polyakov-quark-meson model and compare the
results with those obtained with the pure YM potential. A detailed
discussion is given in Ref.~\cite{Haas:2013qwp}, for related work see
Ref.~\cite{Fukushima:2012qa}.

\section{The glue potential with functional methods in pure gauge
  theory and full QCD}

\begin{figure}[b]
	\centering
	\includegraphics[height=.07\textheight]{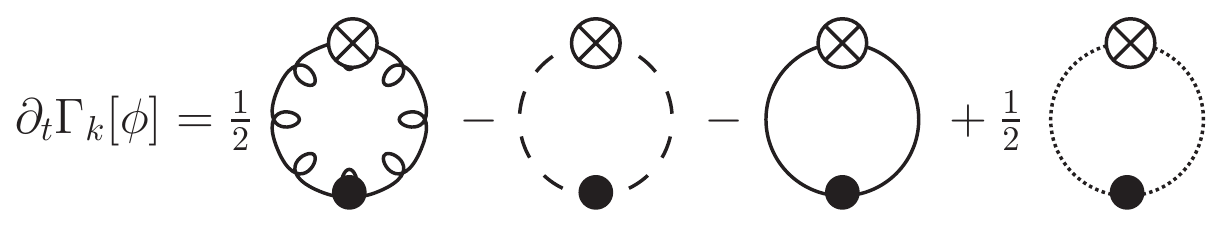}
	\caption{Partially bosonised version of the FRG flow for
          QCD. The loops denote the gluon, ghost, quark and hadronic
          contributions, respectively.  The crosses mark the RG
          regulator term and $t=\log k/\Lambda$ with the infrared
          momentum scale $k$ and the references scale $\Lambda$.}
	\label{fig:QCDflow}
\end{figure}

\begin{figure}[b]
	\centering
	\includegraphics[height=.06\textheight]{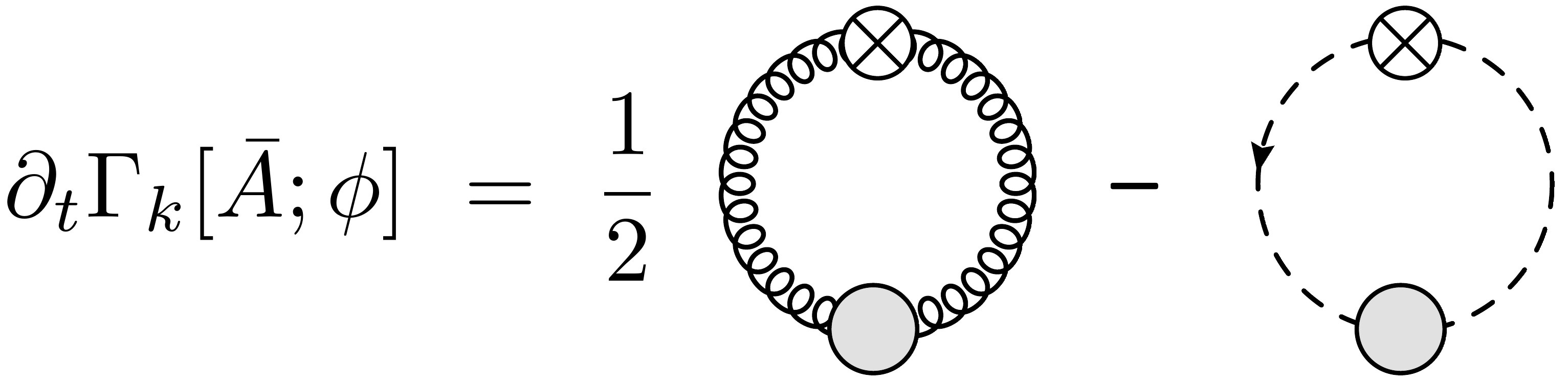}
	\,,\hskip10ex
	\includegraphics[height=.06\textheight]{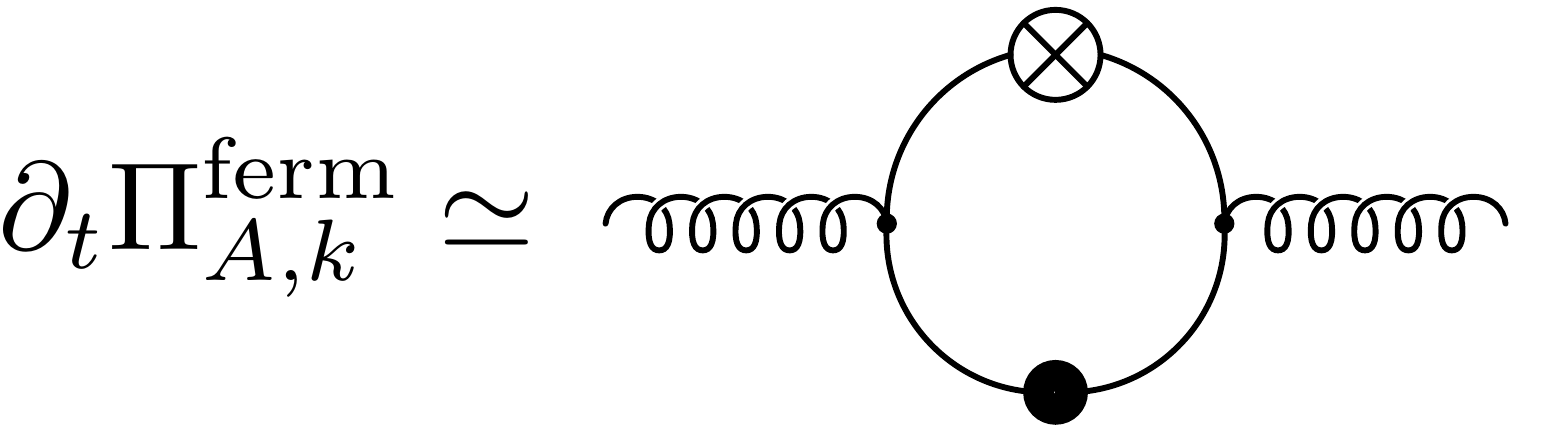}
	\caption{Left: Functional flow for the YM effective
          action. Right: Quark polarisation contribution to the gluon
          propagator representing a contribution of the matter
          backcoupling.  }
	\label{fig:YMflow_quarkloop}
\end{figure}

The functional renormalisation group (FRG) equation for the scale
dependent quantum effective action, the Wetterich equation, is is depicted in Fig.\ref{fig:QCDflow} 
for QCD. 
It provides a setting which allows to approach the low temperature
regime from large momentum scales by successively integrating out
momentum shells with momenta at about $k$.  When lowering the momentum
scale one systematically includes quark and gluon fluctuations into
the theory finally approaching the hadronic phase.  Although the flow
equation has a simple one-loop structure, Fig.~\ref{fig:QCDflow}
describes fully dynamical QCD. In particular, the flow of the gluon
propagator receives contributions from matter loops, e.g.\ the quark
contribution to the vacuum polarisation shown in
Fig.~\ref{fig:YMflow_quarkloop}.

\begin{figure}
	\centering
	\includegraphics[width=.479\textwidth]{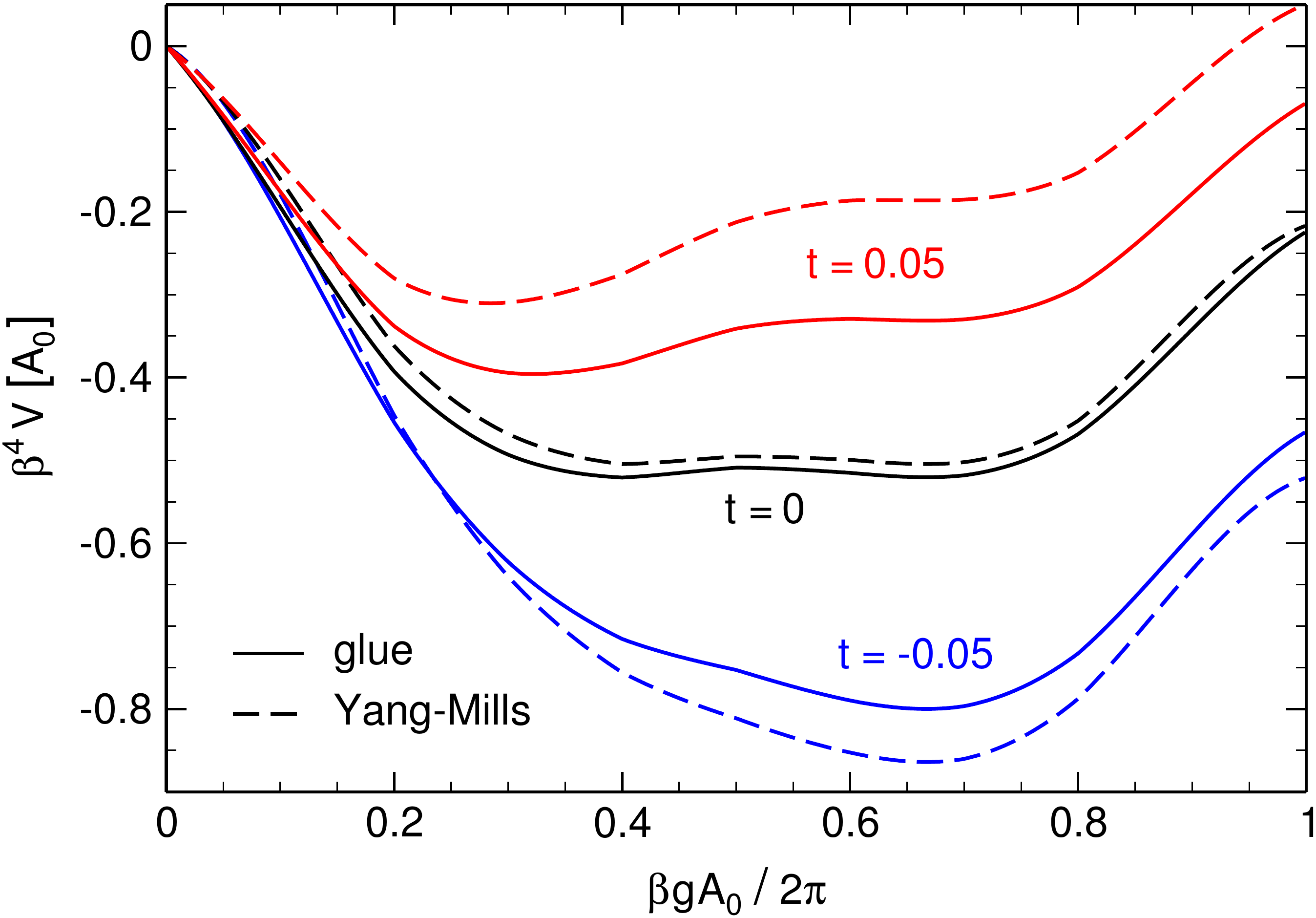}
	\hskip3ex
	\includegraphics[width=.479\textwidth]{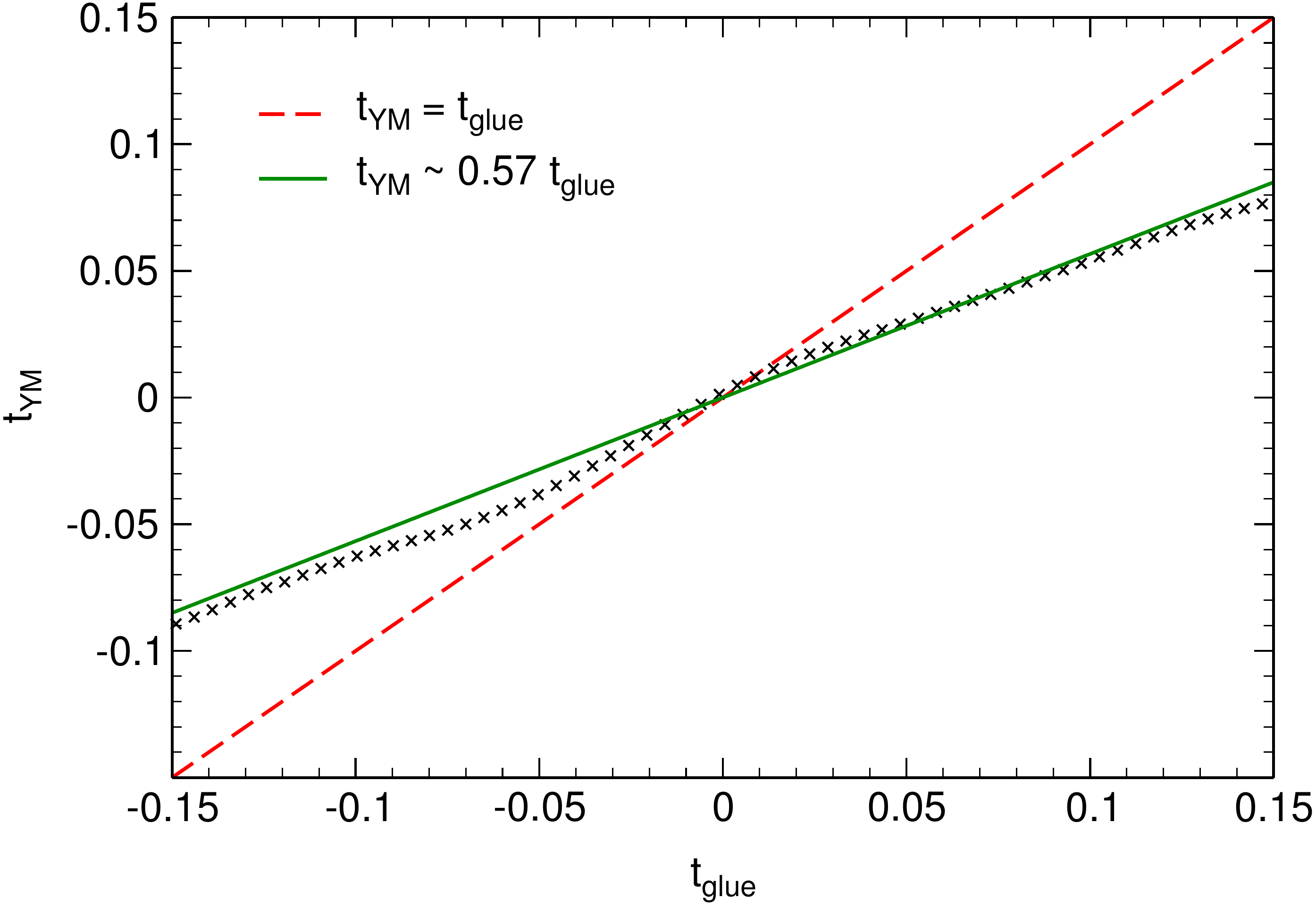}
	\caption{Left: YM potential (dashed lines) and glue effective
          potential (full lines) as functions of the background gauge
          field $\langle A_0 \rangle$ for various reduced
          temperatures. The form of the potentials is similar, however
          the temperature scale changes. Right: Relation between the
          temperature scales of YM theory and the QCD glue potential
          including the backreaction of the quarks. We approximate the
          numerical data (black crosses) by a linear relation (full
          green line) that clearly deviates from pure YM theory (red
          dashed line). }
	\label{fig:YM_glue}
\end{figure}

Evaluated on a general gluonic background $A_0$ and integrated over
all scales $k$, Fig.~\ref{fig:QCDflow} provides the full Polyakov loop
potential of QCD in terms of $A_0$, see Ref.~\cite{Braun:2009gm}. Here the
Polyakov loop $\Phi[A_0]$ serves as the confinement-deconfinement
order parameter \cite{Braun:2007bx,Marhauser:2008fz}. The simple one
loop structure in Fig.~\ref{fig:QCDflow} allows for a straightforward
identification of gluonic and matter contributions. In particular, the
first two loops in Fig.~\ref{fig:QCDflow} generate the glue potential
of QCD and depend on the full gluon and ghost propagators in the
presence of dynamical quarks. For example, for the gluon this entails
the inclusion of the quark part of the vacuum polarisation, see
Fig.~\ref{fig:YMflow_quarkloop}, with fully dynamical quark
propagators.  In turn, dropping the coupling between matter sector and
glue sector, the first two loops in Fig.~\ref{fig:QCDflow} simplify to
the flow of the glue potential of pure YM theory, see
Fig.~\ref{fig:YMflow_quarkloop}.  Within the FRG-approach to two
flavour QCD detailed in Refs.~\cite{Braun:2009gm,Pawlowski:2010ht}, we
have computed the fully dynamical glue potential that takes into
account the backreaction of the quark degrees of freedom on the gluon
propagators (and vice versa).  We compare this potential to the glue
potential of SU$(3)$ YM theory as obtained in
Refs.~\cite{Braun:2007bx,Braun:2010cy,Fister:2013bh}.  The first
non-trivial result of this comparison is that the $A_0$-dependence of
both order-parameter potentials $V\blr{\langle{A_0}\rangle}$ agrees on
a quantitative level, see Fig.~\ref{fig:YM_glue}. This agreement is
achieved after a non-trivial matching of the absolute temperature
scales and the temperature slope of the potentials. Both effects on
the relative temperature dependences are induced by the matter
fluctuations altering the ghost and gluon propagators in QCD.
We exploit these observations to estimate how the
temperature-dependence of a given Polyakov loop model potential for
pure YM theory has to be modified to improve it towards a full QCD
Polyakov loop potential. As a measure for the distance of the
potentials we use \be \int_0^{{2\pi T}/{\bar{g}}} d \langle A_0\rangle
\left |V_{\rm YM}(\langle A_0\rangle) -V_{\rm glue}(\langle
  A_0\rangle)\right|^2\,.
	\label{eq:measure}
\ee
In terms of reduced temperatures 
\be
	t_{\mrm{glue}}=\blr{T-T^{\mrm{glue}}_{\mrm{cr}}}/{
T^{\mrm{glue}}_{\mrm{cr}}} \qquad \mrm{and} \qquad t_{\mrm{YM}}=\blr{T-T^{\mrm{YM}}_{\mrm{cr}}}/{T^{\mrm{YM}}_{\mrm{cr}}}
\ee
the minimisation of the measure (\ref{eq:measure}) yields the
following translation of the two temperature scales
\bea
	t_\mrm{YM} (t_\mrm{glue})&\approx& 0.57\, t_\mrm{glue}\;.
	\label{eq:tYMtglue}
\eea
The critical temperatures of the respective potentials in the FRG
computations are $T^{\mrm{glue}}_{\mrm{cr}}=203\,\mrm{MeV}$
\cite{Braun:2009gm} in two-flavour QCD in the chiral limit and
$T^{\mrm{YM}}_{\mrm{cr}}=276\,\mrm{MeV}$
\cite{Braun:2007bx,Braun:2010cy,Fister:2013bh}. We shall discuss below
that the absolute temperature scale still remains a free parameter
(within a small interval) in our model.  The relation
(\ref{eq:tYMtglue}) can now serve as an import input for model
studies.  For the detailed derivation and further details we refer the
interested reader to Ref.~\cite{Haas:2013qwp}.

\section{Polyakov-quark-meson model and Polyakov-loop potential}

Effective models of QCD such as the PNJL
\cite{Fukushima:2003fw,Ratti:2005jh,Megias:2004hj,Roessner:2006xn} and
the PQM
\cite{Schaefer:2007pw,Schaefer:2009ui,Skokov:2010wb,Herbst:2010rf,Schaefer:2011ex,Mintz:2012mz,Herbst:2013ail}
model unify aspects of chiral symmetry and confinement. The models
are defined via the free energy of QCD as a function of the order
parameters and thermodynamic control parameters
\begin{equation}
	\Omega\blr{\sigma,\sy,\Phi,\Phib;\,T,\mu_f} = U\blr{\sigma,\sy} + 
\mc{U}\blr{\Phi,\Phib;\,T} + \Omega_\qqb\blr{\sigma,\sy,\Phi,\Phib;\,T, \mu_f} \;.
	\label{eq:grand_canon_pot}
\end{equation}
The particle content of the PQM-model is constituent quarks minimally
coupled to gauge fields and coupled to mesons via a Yukawa-type term.
The contribution of the mesonic degrees of freedom $U\blr{\sigma,\sy}$
contains several parameters that are adjusted to meson properties, see
e.g.~Refs.~\cite{Lenaghan:2000ey,Schaefer:2008hk}. The values of the
constants that we use to fix these parameters are given in
Ref.~\cite{Haas:2013qwp}.  A remaining uncertainty in the chiral
sector is the mass of the $\sigma$-meson. Within our model we identify
it with the resonance $f_0\blr{500}$ and we discuss its impact by
varying $m_\sigma$ in the range of $\blr{400-600}\,\mrm{MeV}$.  The
order parameters of chiral symmetry are the light or non-strange
condensate $\sigma$ and the strange chiral condensate $\sy$ that can
be combined to the subtracted chiral condensate $\Delta_\mrm{l,s}$.

For the Polyakov-loop potential $\mc{U}\blr{\Phi,\Phib;\,T}$ a
functional form is chosen that reproduces the temperature dependence
of the Polyakov loop expectation value and the thermodynamics of pure
YM theory as obtained in lattice calculations. We compare results
obtained with the logarithmic parametrisation of
Ref.~\cite{Roessner:2006xn} (that we abbreviate with `Log' in the
following) and the polynomial parametrisation with the parameters of
Ref.~\cite{Scavenius:2002ru} (Poly-I) and those of
Ref.~\cite{Ratti:2005jh} (Poly-II).  We use our relation between the
pure gauge and the glue effective potential (\ref{eq:tYMtglue}) to
effectively convert these pure gauge potentials to the glue potential
of full QCD
\bea
	\label{eq:gluePotYMPot}
	{\mc{U_\mrm{glue}}\blr{\Phi,\Phib,t_\mrm{glue}}}/{T^4} &=&
        {\mc{U_\mrm{YM}}\blr{\Phi,\Phib,t_\mrm{YM}}}/{T_\mrm{YM}^4}
        \;.  \eea
The absolute temperature scale for the case of physical quark masses
(pion masses) has not been computed in
Ref.~\cite{Braun:2009gm}. Therefore, we consider the glue critical
temperature in the model as a free parameter that we vary in the range
$180\,\mrm{MeV} \lesssim T^{\mrm{glue}}_{\mrm{cr}} \lesssim
270\,\mrm{MeV} $ \cite{Schaefer:2007pw}, for details see
Ref.~\cite{Haas:2013qwp}.

The last term of Eq.~(\ref{eq:grand_canon_pot}) represents the quark
sector that includes the coupling to the Polyakov-loop variables and
the mesons.  For a detailed discussion of the 2+1 flavour model we
refer to Refs.~\cite{Haas:2013qwp,Schaefer:2009ui,Mintz:2012mz}.

\section{Results and discussion}

\begin{figure}
	\centering
	\includegraphics[width=.479\textwidth]{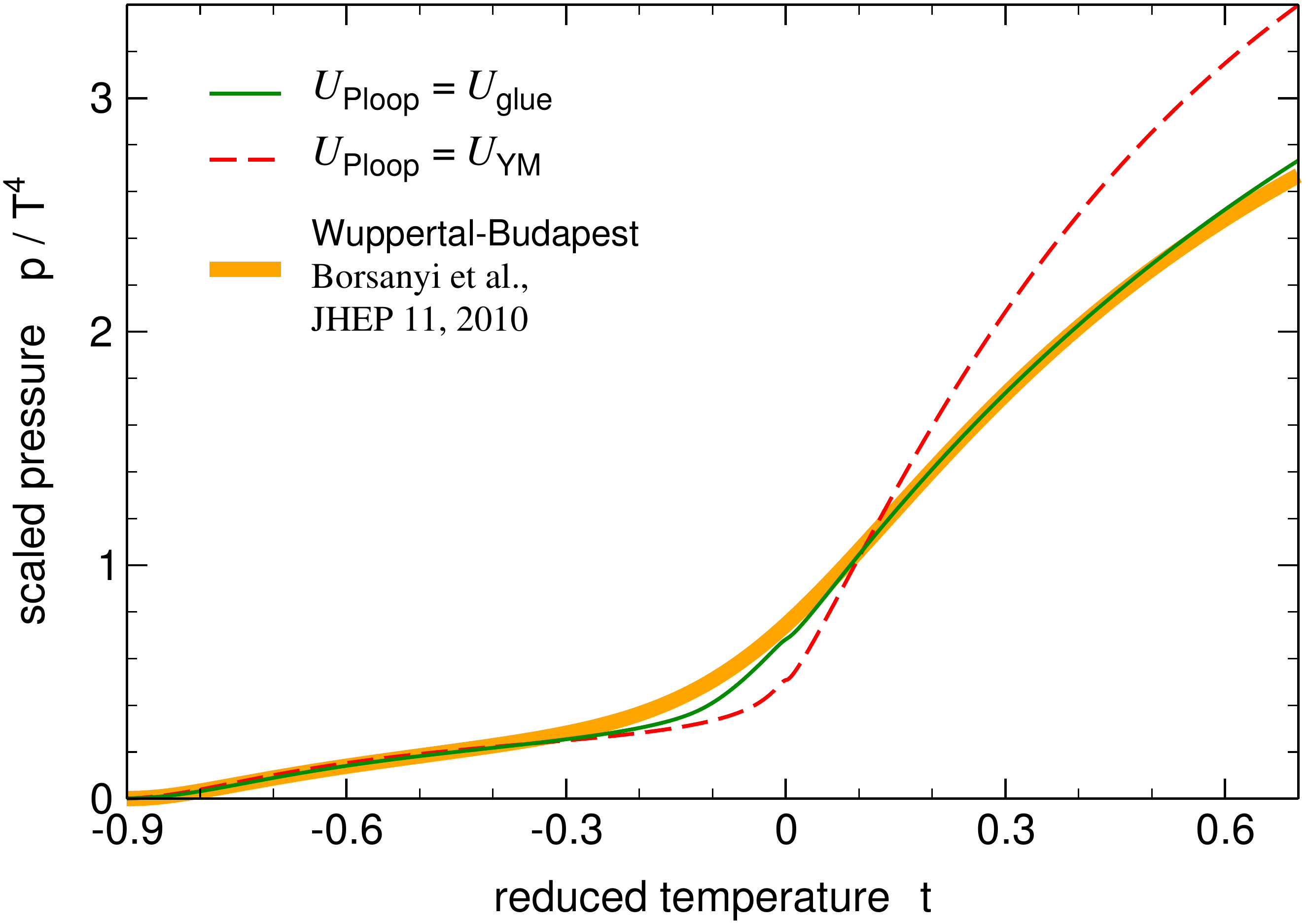}
	\hskip3ex
	\includegraphics[width=.479\textwidth]{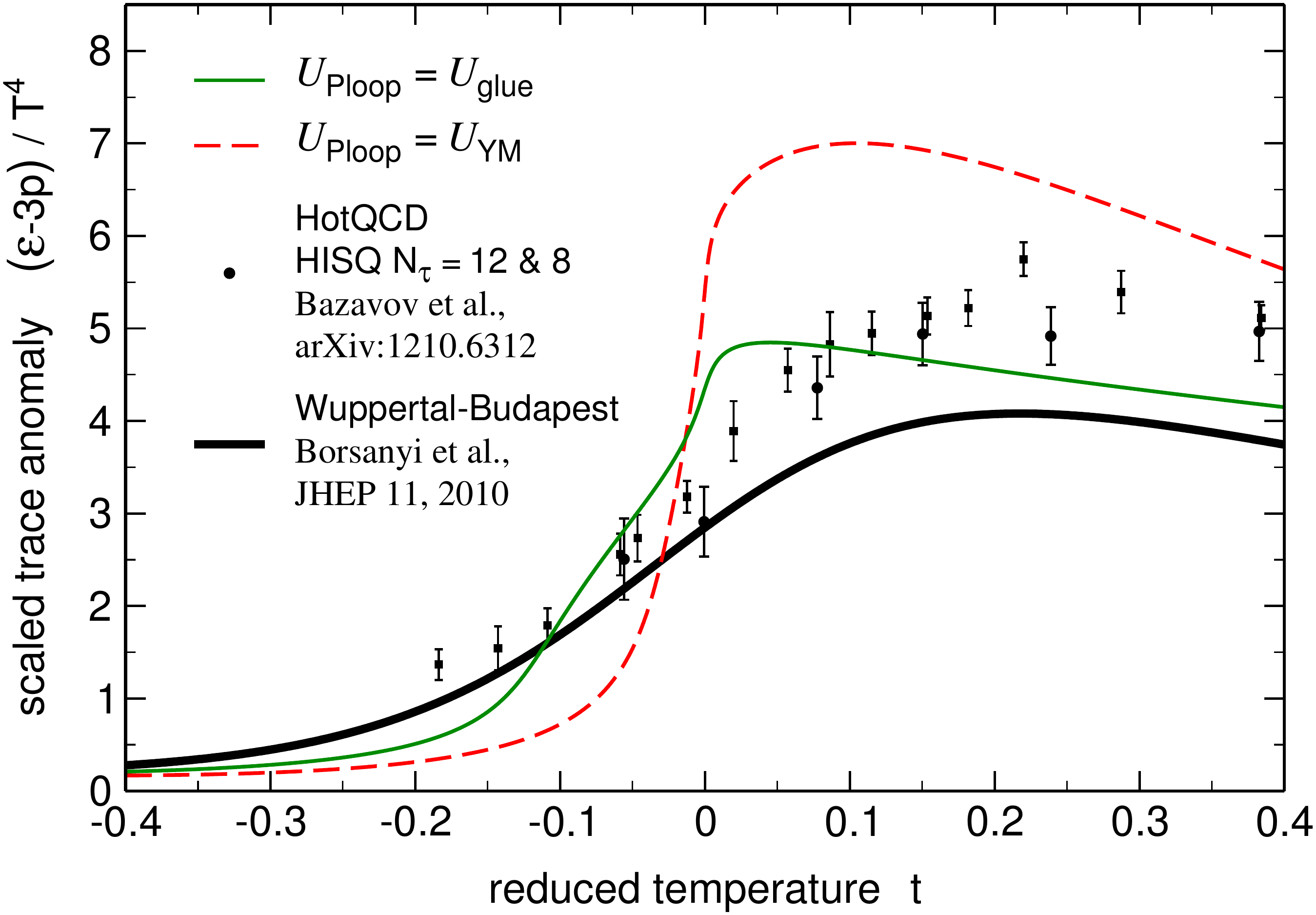}
	\caption{ Normalised pressure (left) and trace anomaly or
          interaction measure (right) approximating the Polyakov-loop
          potential with the pure gauge potential (red dashed lines)
          and with the quark-improved glue potential (full green
          lines), compared to the result of lattice calculations
          \cite{Borsanyi:2010cj, Bazavov:2012bp}, see Table
          \protect\ref{tab:critTemps}.  }
	\label{fig:pe3p_YMglue}
\end{figure}

\begin{figure}
	\centering
	\includegraphics[width=.479\textwidth]{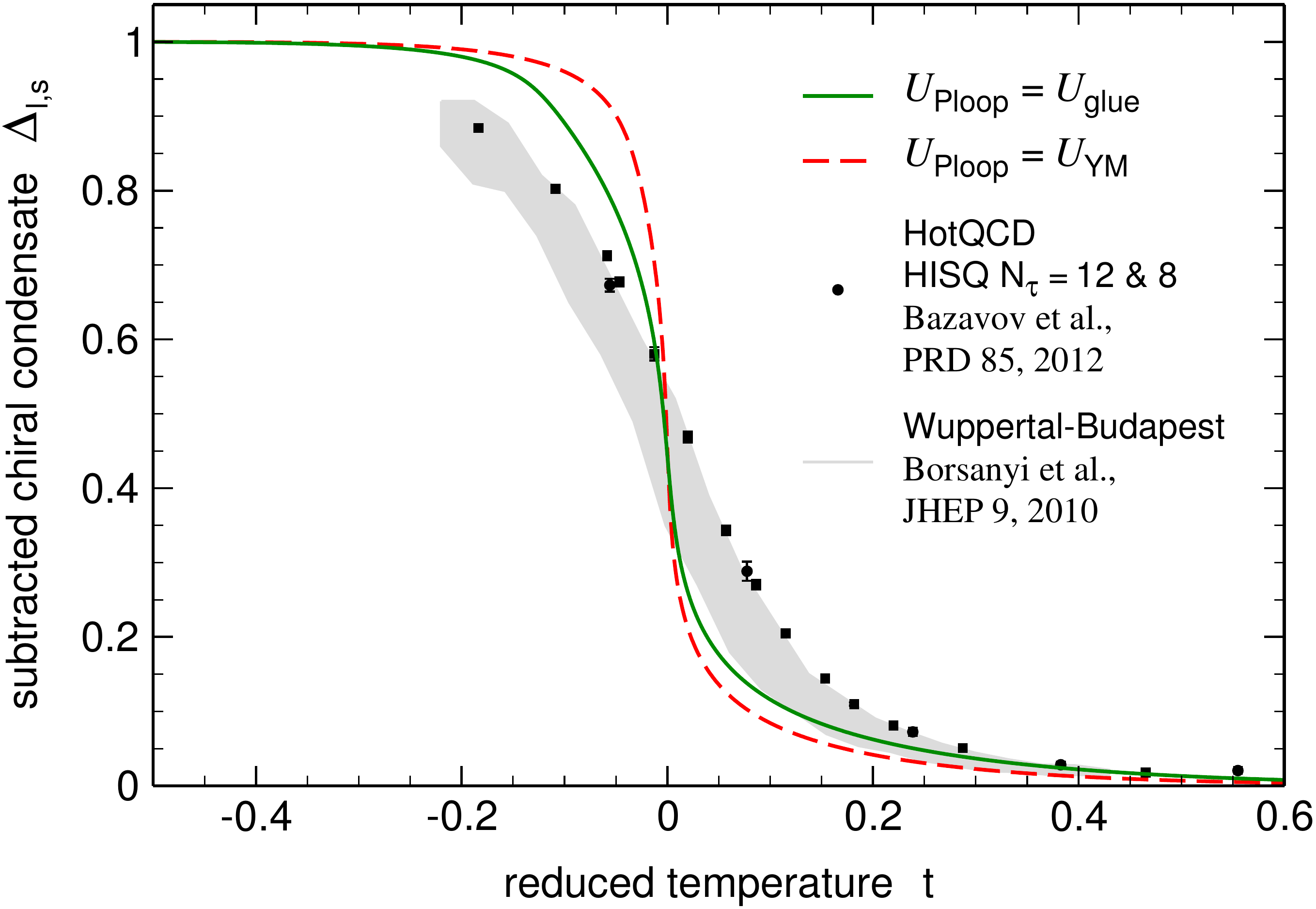}
	\hskip3ex
	\includegraphics[width=.479\textwidth]{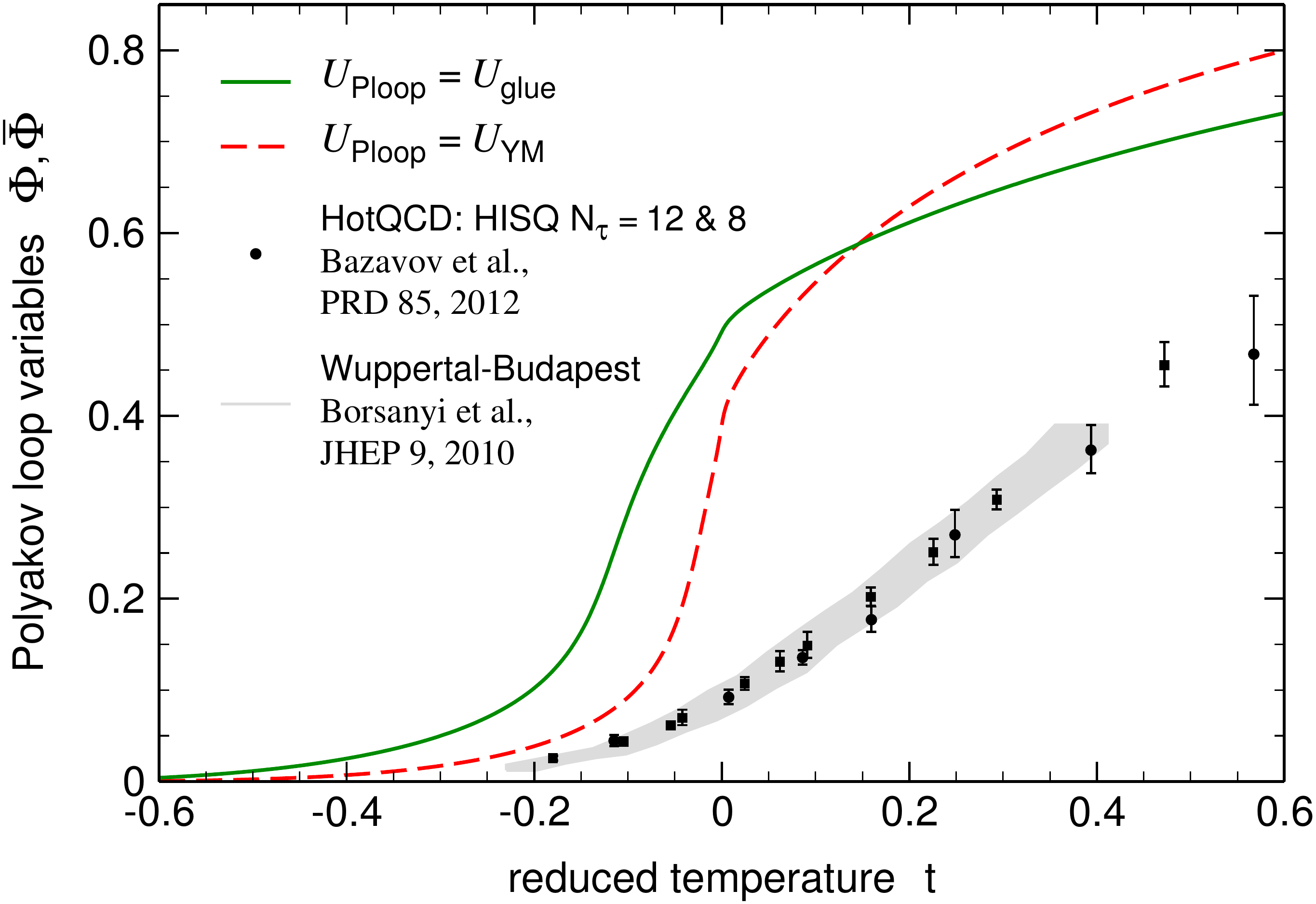}
	\caption{Subtracted chiral condensate (left) and Polyakov loop
          (right) approximating the Polyakov-loop potential with the
          pure gauge potential (red dashed lines) and with the
          quark-improved glue potential (full green lines), compared
          to the result of lattice calculations \cite{Borsanyi:2010bp,
            Bazavov:2011nk}, see also Table
          \protect\ref{tab:critTemps}.  }
	\label{fig:DlsPhi_YMglue}
\end{figure}

The temperature dependence of the thermodynamics and the order
parameters displayed in Figs.~\ref{fig:pe3p_YMglue} and
\ref{fig:DlsPhi_YMglue} shows that replacing the pure YM Polyakov-loop
potential by the quark-improved potential leads to a significantly
smoother crossover transition at vanishing density. The pressure is in
remarkable agreement with the result from lattice calculations given
the mean field nature of the matter sector. The amplitude of the trace
anomaly or interaction measure follows the lattice calculations as
well. We observe that the behaviour of the pressure and interaction measure in the transition region
is still slightly steeper.
Also the evolution of the chiral order parameter shows a smoother
evolution when applying the quark-improved Polyakov-loop potential.
Including fluctuations will bring the smoothening towards the
lattice result as will be discussed in a forthcoming publication.
The pseudocritical temperatures summarised in Table
\ref{tab:critTemps} show that the result with the quark-improved
Polyakov-loop potential agrees well with the value found in lattice
calculations.
\begin{table}
	\centering
	\begin{tabular}{l|c|c|c|c}
					& $\mc{U}_\mrm{YM}$	& $\mc{U}_\mrm{glue}$	& Wuppertal-Budapest \cite{Borsanyi:2010bp}		& HotQCD $N_\tau=12$ \& $8$ \cite{Bazavov:2011nk} \\ \hline
		$T_\mrm{c}$ [MeV]	& 168			& 158			& 157				& 159 \& 163
	\end{tabular}
	\caption{Pseudocritical temperatures for the crossover phase transition at $\mu_f=0$. They are determined by the peaks in the temperature derivatives of the subtracted condensate $\Delta_\mrm{l,s}$. For the effective model calculations we use the logarithmic parametrisation of the Polyakov-loop potential with a critical temperature of $210\,\mrm{MeV}$ and a sigma meson mass of $500\,\mrm{MeV}$.}
	\label{tab:critTemps}
\end{table}
The result for the Polyakov loop is shifted to higher values in the
confined phase and to lower values in the deconfined phase when
replacing the YM Polyakov-loop potential by the effective glue
potential, smoothening the transition. This is a consequence of
applying Eq.~(\ref{eq:tYMtglue}) as can be also seen in
Fig.~\ref{fig:YM_glue}.  The remaining difference can be attributed to
the missing quark-meson dynamics and that the Polyakov loop potential
is written in terms of the variable $\langle\Phi[A_0]\rangle$ while
the coupling to the matter sector is described as a function of
$\Phi\left[\langle{A_0}\rangle\right]$, see the discussion in
Refs.~\cite{Haas:2013qwp,Pawlowski:2010ht,Braun:2011fw}.

The uncertainties within our model study are the parametrisation and
the transition temperature of the Polyakov-loop potential and the mass
of the $\sigma$-meson.
\begin{figure}
	\centering
	\includegraphics[width=.479\textwidth]{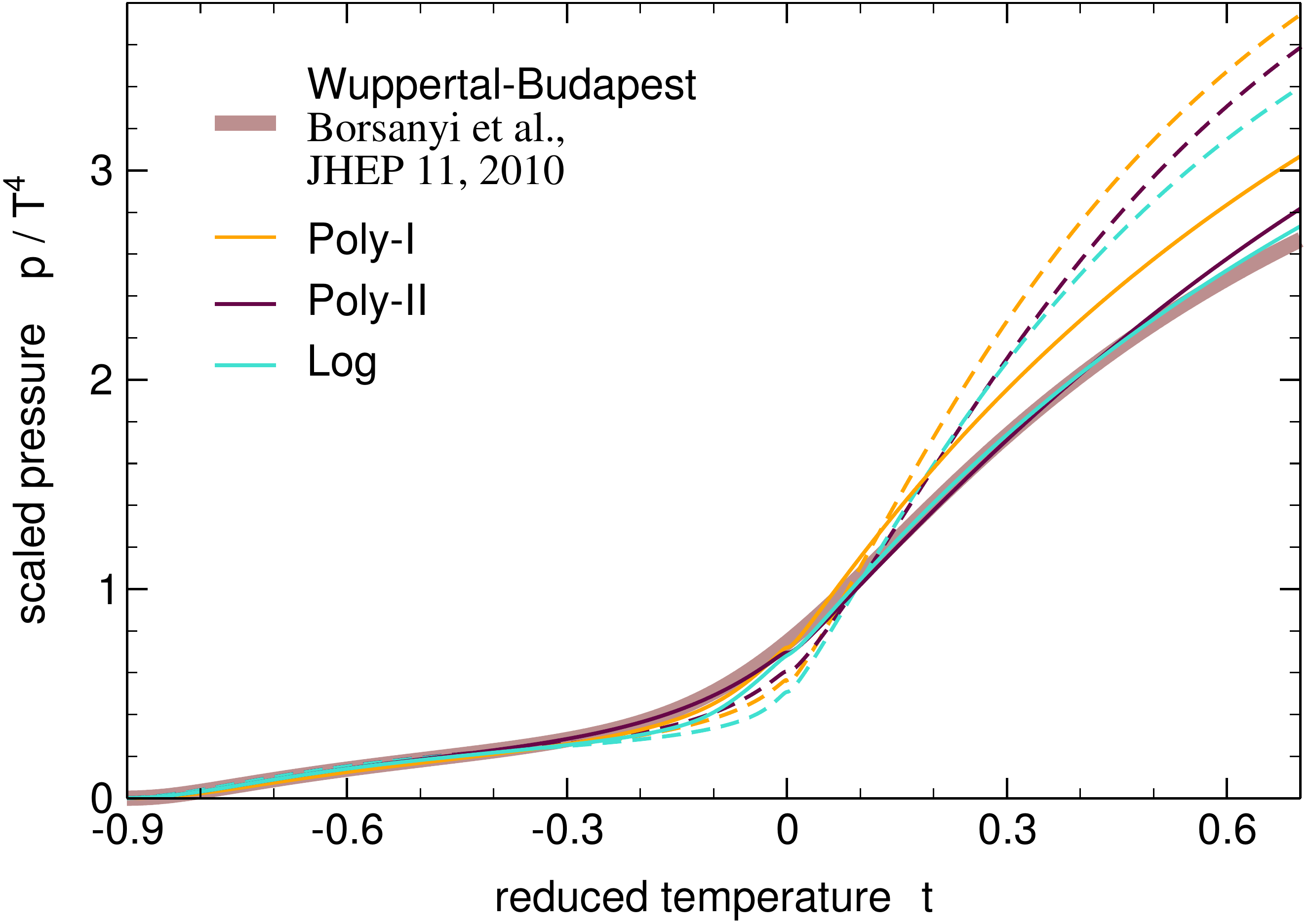}
	\hskip3ex
	\includegraphics[width=.479\textwidth]{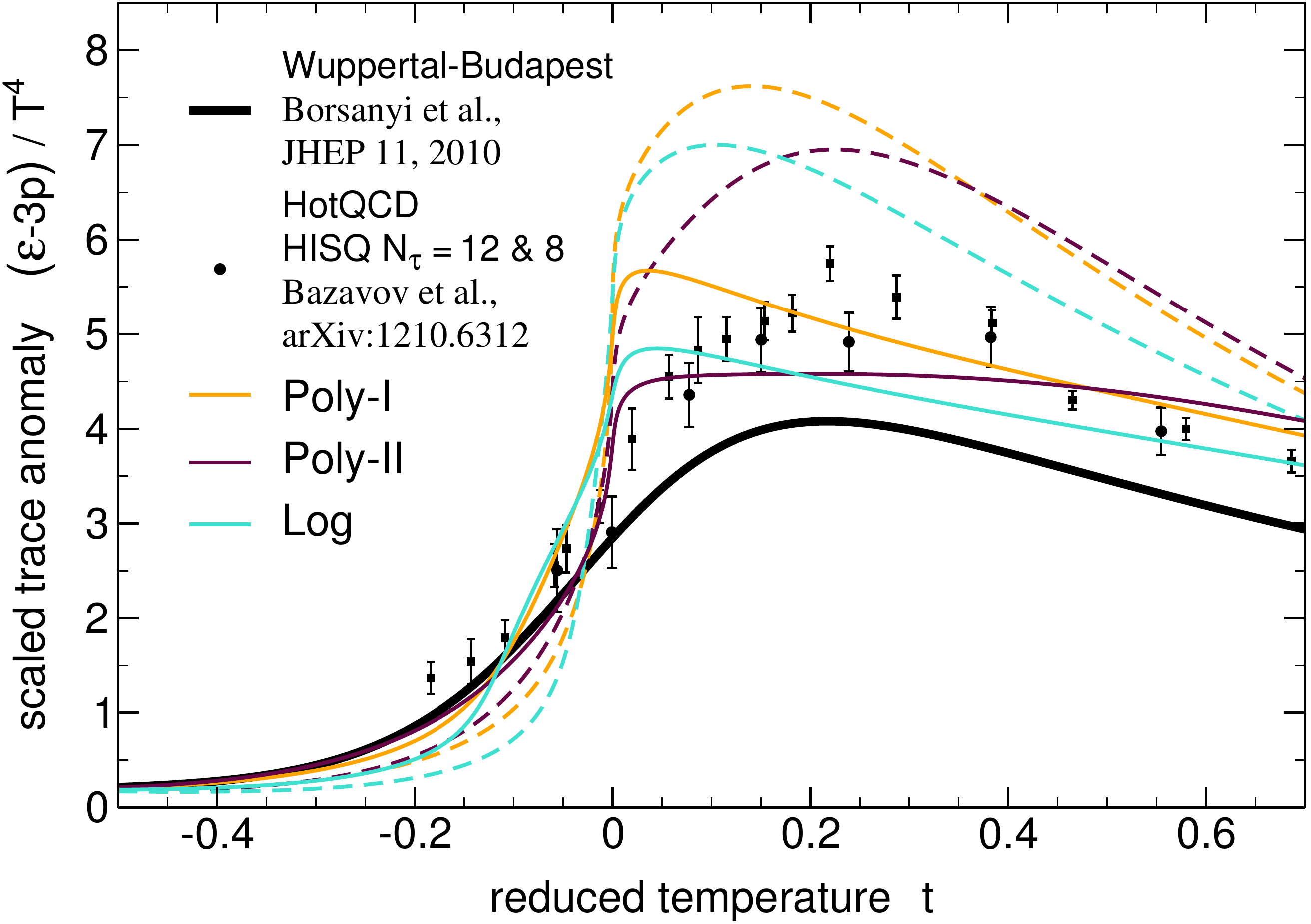}
	\caption{Normalised pressure (left) and trace anomaly or
          interaction measure (right) for different parametrisations
          of the Polyakov-loop potential and compared to the result of
          lattice calculations \cite{Borsanyi:2010cj,
            Bazavov:2012bp}. Dashed lines stand for the case of
          approximating the Polyakov-loop potential with the YM
          potential and solid lines are for the case of using the
          improved glue potential. We use a transition temperature of
          the Polyakov-loop potential of $210\,\mrm{MeV}$ and a sigma
          meson mass of $500\,\mrm{MeV}$.}
	\label{fig:pe3p_PloopPots}
\end{figure}
Figure~\ref{fig:pe3p_PloopPots} shows the sensitivity on the
parametrisation and different parameter sets of the glue sector. The
results within the different descriptions of the YM or glue potential
are compatible but the transition is in general smoother with the
quark-improved glue potential. The amplitude of the trace anomaly is
in better agreement with lattice calculations. Note, however, that the
temperature dependence is slightly steeper independent of the actual
parametrisation. The maximum of the interaction measure above
$T_\mrm{c}$ with the Poly-II YM Polyakov-loop potential corresponds to
the nearly constant behaviour above $T_\mrm{c}$ when the glue
Polyakov-loop potential is applied.  Table
\ref{tab:critTemps_Params_T0s} summarises the pseudocritical
temperatures for the different calculations. Both the steepness as
well as the flat behaviour above $T_c$ are expected shortcomings of
the present mean-field nature of the matter dynamics. The effects of
including a fully dynamical matter sector will be discussed in a
forthcoming publication.
\begin{table}
	\centering
	\begin{tabular}{l|c|c|c}
					& Poly-I	& Poly-II	& Log \\ \hline
		$\mc{U}_\mrm{YM}$	& 158 		& 169 		& 168 \\
		$\mc{U}_\mrm{glue}$	& 144 		& 158	 	& 158
	\end{tabular}
	\qquad\qquad
	\begin{tabular}{c|c|c|c|c}
			$T^{\mrm{glue}}_\mrm{cr}$ [MeV]	& 180 & 210 & 240 & 270 \\ \hline
			$\mc{U}_\mrm{YM}$		& 156 & 168 & 183 & 198 \\
			$\mc{U}_\mrm{glue}$		& 152 & 158 & 164 & 171 
	\end{tabular}
	\caption{Pseudocritical temperatures in MeV for the crossover transition at $\mu_f=0$. Left: For different parametrisations and parameter sets of the Polyakov-loop potential using a transition temperature of the potential of $210\,\mrm{MeV}$. Right: For different critical temperatures of the Polyakov-loop potential using the logarithmic parametrisation. For both cases we use a sigma meson mass of $500\,\mrm{MeV}$. The pseudocritical temperatures are determined by the peaks of the chiral susceptibility $\pd{\Delta_\mrm{l,s}}/\pd{T}$.}
	\label{tab:critTemps_Params_T0s}
\end{table}
\begin{figure}
	\centering
	\includegraphics[width=.479\textwidth]{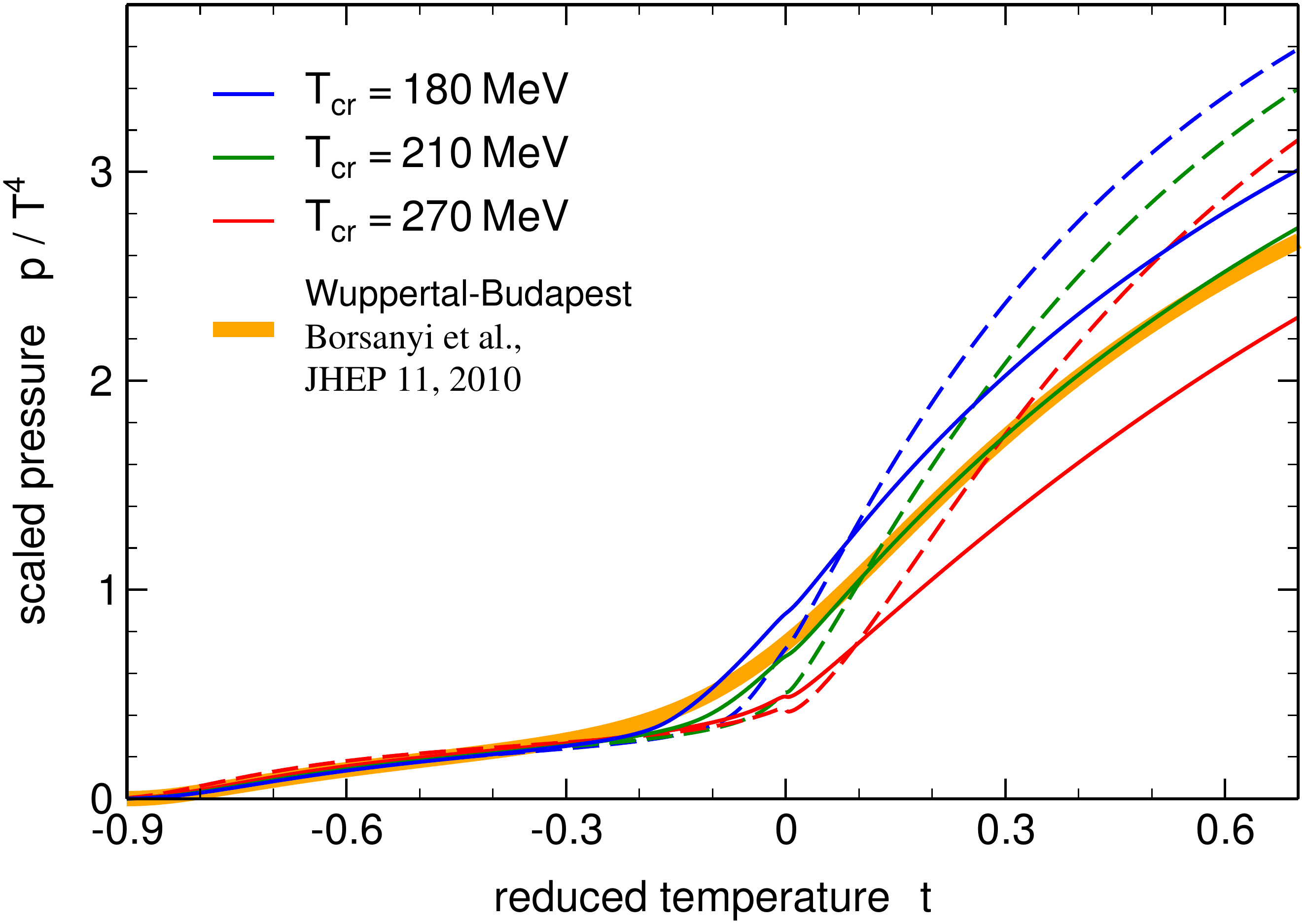}
	\hskip3ex
	\includegraphics[width=.479\textwidth]{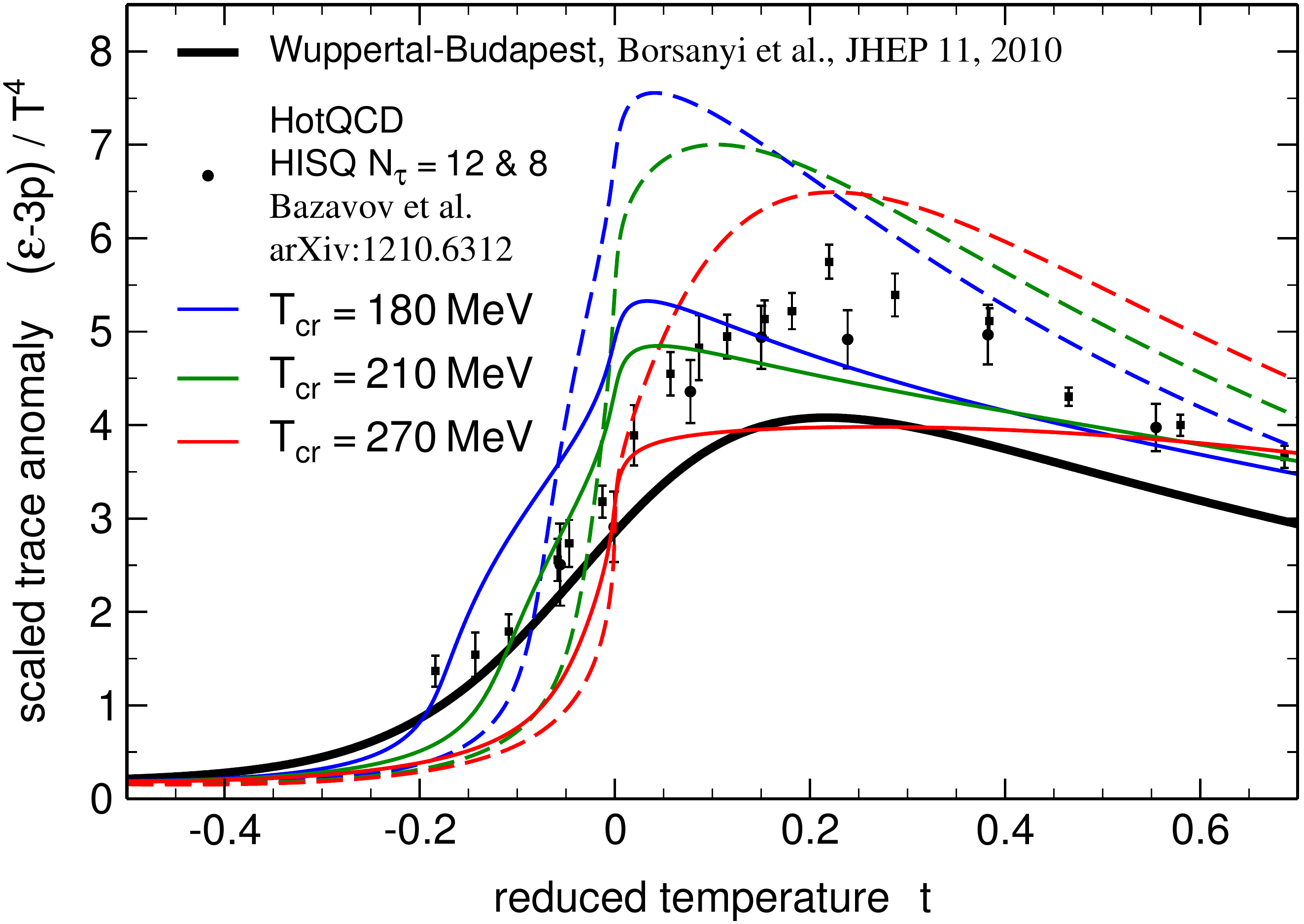}
	\caption{Normalised pressure (left) and trace anomaly or
          interaction measure (right) for different transition
          temperatures of the Polyakov-loop potential compared to the
          result of lattice calculations \cite{Borsanyi:2010cj,
            Bazavov:2012bp}. Dashed lines are for the YM potential
          case and solid lines are for the improved glue
          potential. The logarithmic parametrisation of the
          Polyakov-loop potential and a sigma meson mass of
          $500\,\mrm{MeV}$ is used.}
	\label{fig:pe3p_T0s}
\end{figure}

Our analysis of the impact of the transition temperature $T_{\rm
  cr}^{\rm glue}$ of the Polyakov-loop potential on thermodynamic
observables shown in Fig.~\ref{fig:pe3p_T0s} confirms that the
temperature dependence of the pressure and the interaction measure is
in far better agreement with lattice calculations when the
quark-improved Polyakov-loop potential is used. The transition gets
steeper when the transition temperature of the glue potential is
decreased. We find best agreement with lattice calculations for
$T_\mrm{cr}^\mrm{glue}\sim210\,\mrm{MeV}$ but this scale could be
lowered by the inclusion of fluctuations \cite{Herbst:2013ail}. The
temperature dependence of the trace anomaly shows that the transition
region expands below the pseudocritical temperature if the transition
temperature $T_{\rm cr}^{\rm glue}$ of the glue potential is
decreased. The pseudocritical temperature is at the lower end of the
transition region for larger critical temperatures of the
Polyakov-loop potential.  The absolute values of the pseudocritical
temperatures are given in Table \ref{tab:critTemps_Params_T0s}.

\begin{figure}
	\centering
	\includegraphics[width=.479\textwidth]{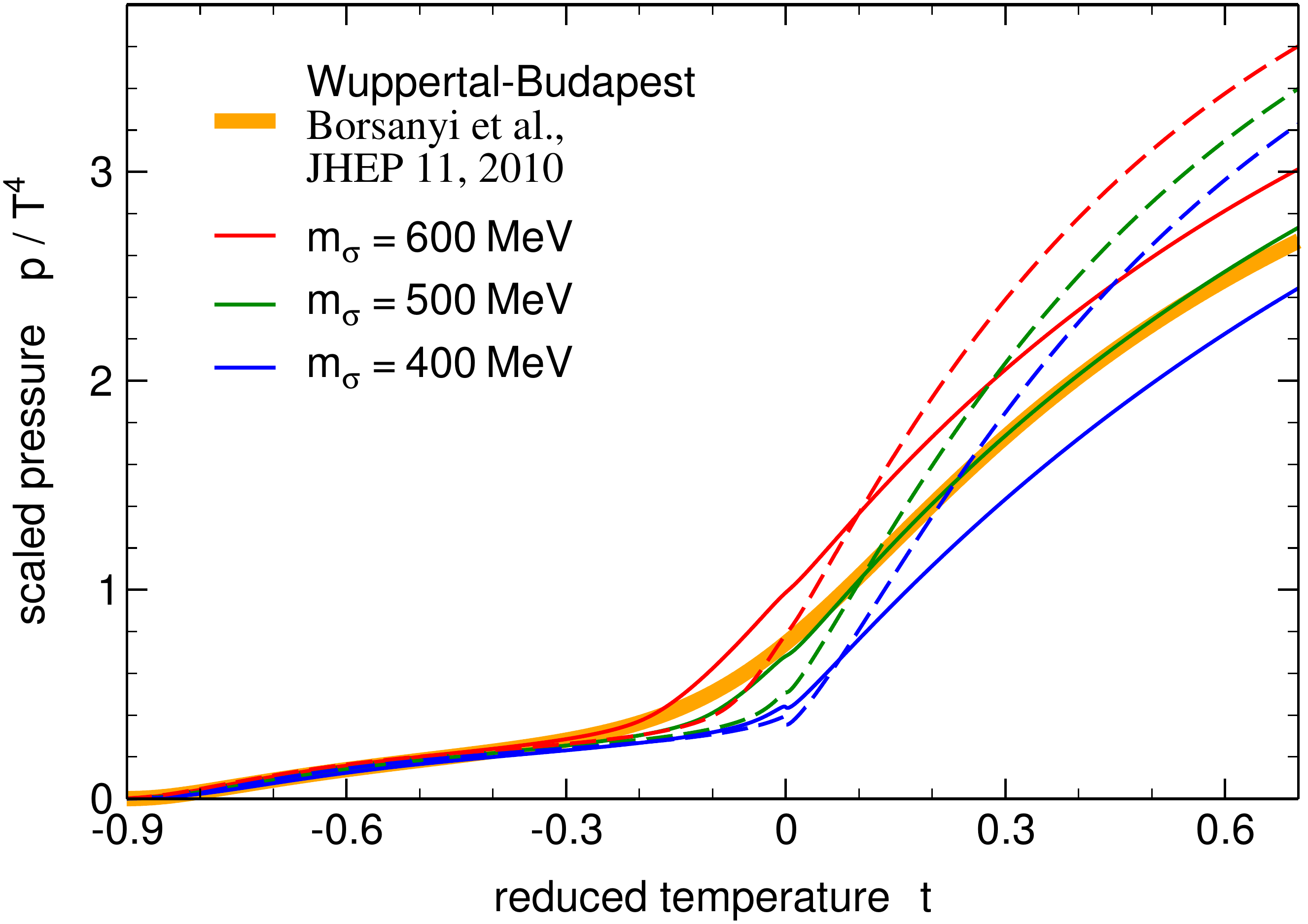}
	\hskip3ex
	\includegraphics[width=.479\textwidth]{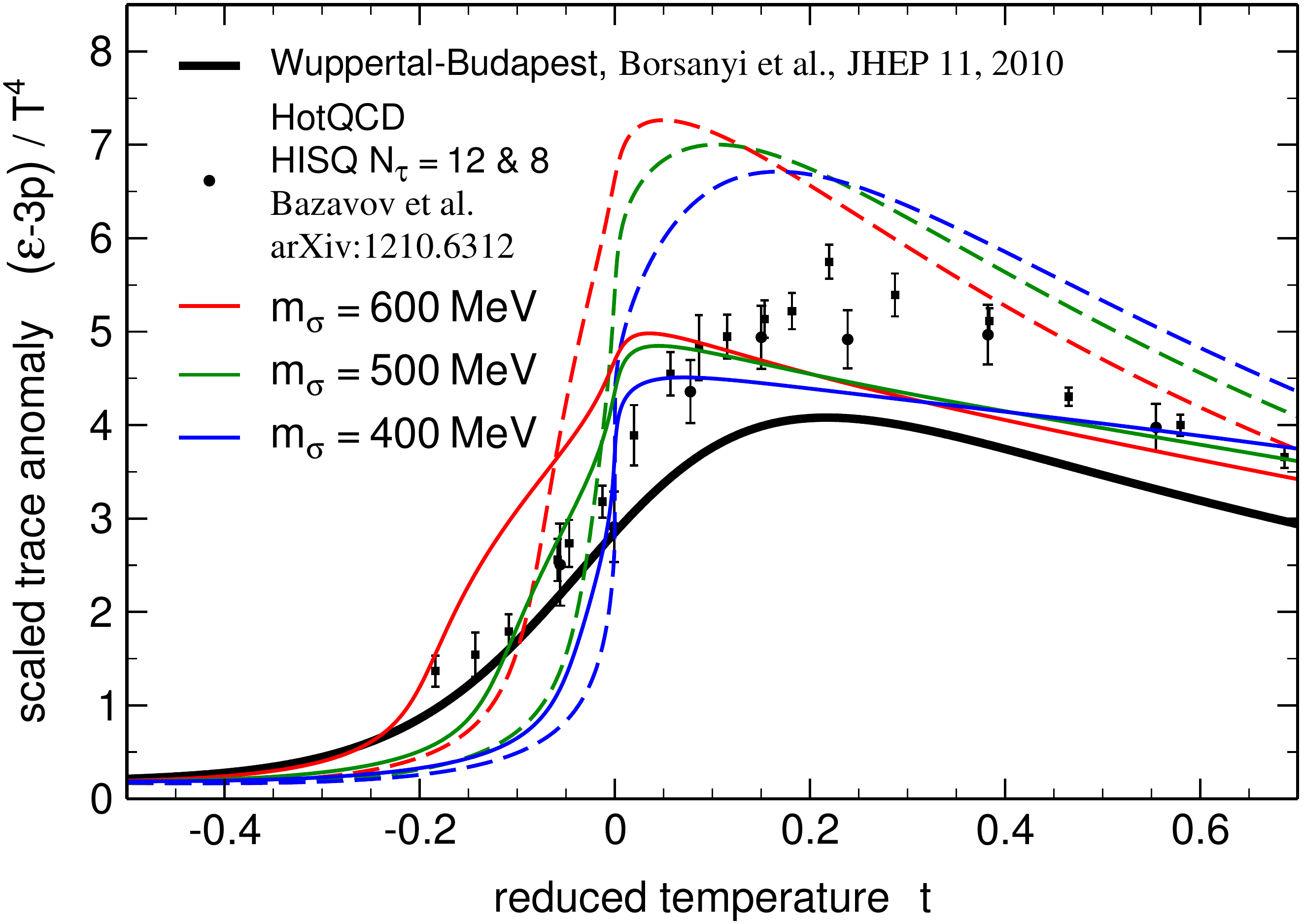}
\caption{Same as the previous figure for different masses
          of the scalar sigma meson. The logarithmic parametrisation
          of the Polyakov-loop potential and a critical temperature of
          $210\,{\rm MeV}$ is used.}
	\label{fig:pe3p_msigmas}
\end{figure}
The dependence of the results on the mass of the $\sigma$-meson
(Fig.~\ref{fig:pe3p_msigmas}) shows that the pseudo\-critical
temperature moves from the lower end of the transition region to the
upper end when the $\sigma$-meson mass is increased. Increasing
$m_\sigma$ has an opposite effect to increasing the transition
temperature of the Polyakov-loop potential. The quantitative best
agreement with the pressure of lattice calculations is obtained for
the combinations $(m_{\sigma}, T_{\rm cr}^{\rm glue})$ of the
$\sigma$-meson mass and the Polyakov-loop potential transition
temperature of (400\,MeV, 180\,MeV), (500\,MeV, 210\,MeV) and
(600\,MeV, 250\,MeV). The best agreement with lattice QCD data for the
pseudocritical temperature is found for the combination
$m_\sigma=500\,\mrm{MeV}$ and $T_\mrm{cr}^\mrm{glue}=210\,\mrm{MeV}$,
as can be read off from Table \ref{tab:critTemps_msigma}.
\begin{table}
	\centering
	\begin{tabular}{c|c|c|c}
		$m_\sigma$ [MeV]	& 400 & 500 & 600 \\ \hline
		$\mc{U}_\mrm{YM}$ 	& 161 & 168 & 179 \\
		$\mc{U}_\mrm{glue}$	& 144 & 158 & 173
	\end{tabular}
	\caption{Pseudocritical temperatures for the crossover transition at $\mu_f=0$ for different masses of the sigma meson. They are determined by the peaks in the temperature derivatives of the subtracted condensate $\Delta_\mrm{l,s}$. We use the logarithmic parametrisation of the Polyakov-loop potential with a critical temperature of $210\,\mrm{MeV}$.}
	\label{tab:critTemps_msigma}
\end{table}

\acknowledgments We thank L.~Fister, E.~S.~Fraga, T.~K.~Herbst,
B.~W.~Mintz and B.-J.~Schaefer for discussions and collaboration on
related topics. This work is supported by BMBF under grants FKZ
05P12VHCTG and 06HD7142, by ERC-AdG-290623, by the Helmholtz Alliance
HA216/EMMI, by the DFG through the HGSFP, by HGS-HIRe, and by HIC for
FAIR within the LOEWE program.

\bibliography{proceed_refs}

\end{document}